# Finger-Stylus for Non Touch-Enable Systems

Ankit Chaudhary

Member IEEE
Independent Researcher
Iowa City, IA USA
dr.ankit@ieee.org

*Abstract*—Since computer was invented, people are using many devices to interact with computer. Initially there were keyboard, mouse etc. but with the advancement of technology, new ways are being discovered that are quite usual and natural to the humans like stylus, touch-enable systems. In the current age of technology, user is expected to touch the machine interface to give input. Hand gesture is such a way to interact with machines where natural bare hand is used to communicate without touching machine interface. It gives a feeling to user that he is interacting in natural way to some human, not with traditional machines. This paper presents a technique where user needs not to touch the machine interface to draw on screen. Here hand finger draws shapes on monitor like stylus, without touching the monitor. This method can be used in many applications including games. The finger was used as an input device that acts like paint-brush or finger-stylus and is used to make shapes in front of the camera. Fingertip extraction and motion tracking were done in Matlab with real time constraints. This work is an early attempt to replace stylus with the natural finger without touching screen.

## I. INTRODUCTION

Communication using hand gestures exist in all civilizations since old time. There is a specific way to present hand to show a particular message. A gesture is a form of non-verbal communication in which any message is conveyed with the help of visible body actions. Hand gesture is potentially a very natural and useful modality for human-machine interaction. Hand gesture recognition (HGR) now has been an adoptable and reliable way to communicate with machines [1]. People are using it to control robots [2], to learn/interpret sign languages [3-6], in health care [7] and many other fields. The use of hand gesture, as an interface between human and machines has always been a very attractive alternative to the conventional interface devices. HGR has been applied to many applications using different techniques since last three decades. Till date, mostly it is sensors or touch based recognition on devices which is used for interaction. The sensor-glove based method hinders the ease and naturalness with which humans interact with the computer. This has led to an increased interest in the visual approach.

These days tablets and touch enabled mobiles are in use and mostly users go for these devices like Apple iPad, Samsung Galaxy etc. They have few applications (apps) which are controlled or run by touching the screen in normal way or in a specific way. For example minimization of apps, entering data, paint. It may be single finger gesture or multi finger gesture. Here to start with, we are focusing on one application which is paint. It is like 'Paper' or 'Fingerpaint Magic' available for iPad. They work same as 'Paint' works on Microsoft Windows. There may be many applications like these on different platforms. A detailed analysis of paint with different computer vision algorithms has been done by Booch [8]. Forsline [9] came with stylus to use a pen kind e-stick to draw or write on computer screen which changes the world of interaction with computers. *Sensu* is the brush with stylus which works on these devices and gives a feeling to paint on canvas [10]. After this, sensors detecting human body touch on screen changed the technology of interaction. It seems very natural to draw or point on machines using sensors for example interaction with iPad.

In touch enable machines, apps developed on different platforms provide flexibility to draw anything on computer canvas (which is screen) using hand fingers. Even few applications use finger pressure to decide the thickness of paint brush. Among different parts of the body, hand is the easiest to use and show expression of human feelings. Also it is very robust in its operations because of its design and can move in any direction. A good comparison is given in [11] between stylus and hand touch devices. Here we are presenting a method to perform same action by not touching screen. The paper discusses the implementation of paint drawing on computer screen in real time using vision based method, where touch-enable device is not required. Here hand finger works as they stylus, say *finger-stylus*. This can be used on different tablets and replace many exiting apps as discussed above because of its easy to usage and mostly all tablets have camera.

## II. BACKGROUND

The robust tracking of hand has been an active area of research in the applications where finger movements or hand geometry detection is needed. The existing methods are generally divided into two categories: Vision based approach [12] and Glove based approach [13]. Both of them have their own pros and cons. Rehg [14] presented a method to detect articulated hand motion. He proposed 27 degree of freedom of hand in grey images but his method was not effective with complex backgrounds. Sato [15] used infrared cameras for skin segmentation on a table top and template matching for interpreting gesture commands. Chaudhary [2] have also developed a real time finger motion detection framework which can be applied to many applications.

Garry [16] used virtual environments with gestures to control it in natural way. Zeller [17] also used hand gestures in virtual environments. Starner [4] developed a system with single color camera to track American sign language in real time. Bragatto [5] translated Brazilian sign language from video by tracking hand gestures. He used perceptron ANN for color segmentation and then classification separately. Cooper [6] presents a method to control a complex set of sign language with *viseme* representation to increase the lexicon size. Ju [18] used hand gestures to analysing and annoting video sequences of tech talks. In his work, gestures like pointing and writing were detected and recognized. Even systems which can detect the hand pointing spot has been developed [19].

Araga [20] presented a real time gesture recognition system from video where they used few gesture images as gesture states. Wachs [21] developed a gesture based tool for sterile browsing of radiology images. A robust object detection method in indoor and outdoor field is described by Wang [22]. The hand gesture could replace stylus on touch screens or touch screen sensors which are currently used in many places. A similar gesture based patient care system is described in [7]. Coral Inc. has developed fingertip paint brush for touch enable systems [23]. Hettiarchchi [24] presented *FingerDraw* where children can paint on interactive screen with 'worm finger' device.

## III. SYSTEM DESCRIPTION

The system uses a webcam either in-built in system or plugged. It tracks the hand movements made by the user by detecting the finger tips. These tips are displayed on the screen and finally the system shows the whole movement made by the user on the screen by connecting them. This movement further can be used to interact with the computers, mobile devices without any physical intervention. For the simplicity of the system and to keep it near to human behaviour, we are considering only one finger for the tracking and display as one brush on canvas. Later multiple criteria can be added to make the system for more applications.

   Generally human uses index finger to point or to show gesture in their daily life. A 'session' is the time when finger come in front of camera view and goes out of its capturing frame. If the user again bringing a finger in capturing zone, the system will again start its display. User must show only one finger (any finger) as shown in figure 1.

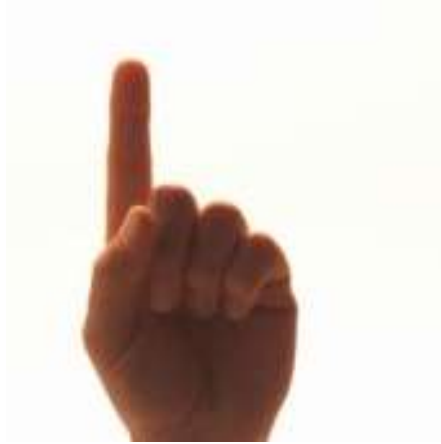

Figure 1: Finger captured by camera

   The system is implemented in Matlab with considering real time constraints. The user is free to move its hand in whatever direction and whichever shape he wants to draw in the air, virtual drawing. Hand image is captured in RGB color space and $YC_bC_r$ color space based skin filter was applied to minimize the color variations. Also segmentation is working robustly irrespectively of light intensity chagnes. The pre-processing results are shown in figure 2.

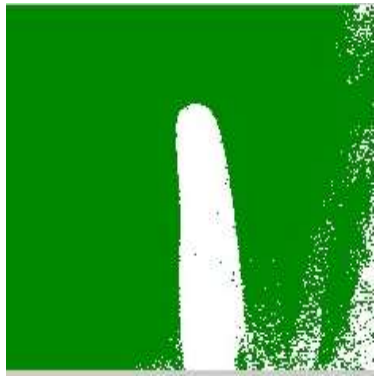

Fig 2: Pre-processing after skin filter

   Different fingertips detection method was tested to check real time response including curvature based, edge based and finally we came with a faster method which is described in [25]. The results are shown in figure 3. This method is direction invariant and fastens the pre-processing by cropping the image by a factor more than 2 depending on the skin pixels in the captured image.

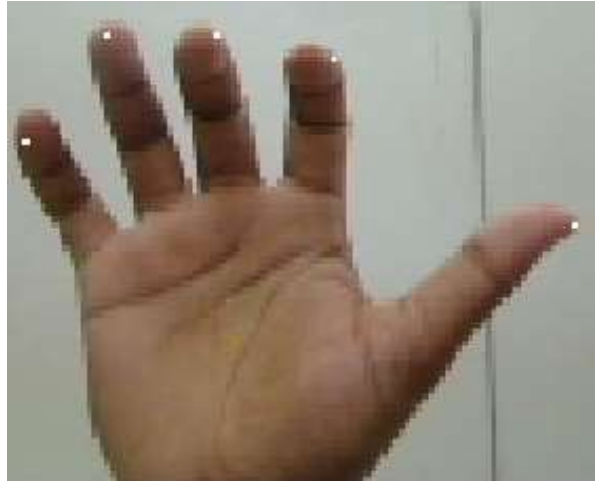

Figure 3: fingertips detection results

## IV.   IMPLEMENTATION

As the system usage, the user should show one finger to the system. If he shows more, the first finger from the left would be chosen. Also it is advisable that finger should point to upward though developed method is able to detect fingertip in any direction. The skin filter was applied on captured image frame and one finger was chosen to consider. Fingertip detection was done by applying an increasing constant to all pixels to the image from wrist to fingertip. The constant ranges from 0 to 255 starting from wrist. Where pixels have value as 255, it would be detected as fingertips.

Mathematically the process can be defined as:

$$Finger_{edge}(x, y) = \begin{cases} 1 & if\ modifiedimage\ (x,y) = 255 \\ 0 & otherwise \end{cases}$$

As we are taking only fingertip pixels which are on the boundary of finger, so other pixels or noise is very less to occur and we would get a sharp fingertip. Also a finger template matching was also done to get first finger from the captured image, this also help in noise reduction. To increase the width of fingertips -5 to +5 pixels are considered as fingertip and marked as red color. This would be thickness of line in drawing. It needs to be changed, if someone needs different thickness in the drawing.

The implementation supports real-time responses and gives a feeling of finger paint brush. The detected fingertips in real time on computer screen are shown in figure 4.

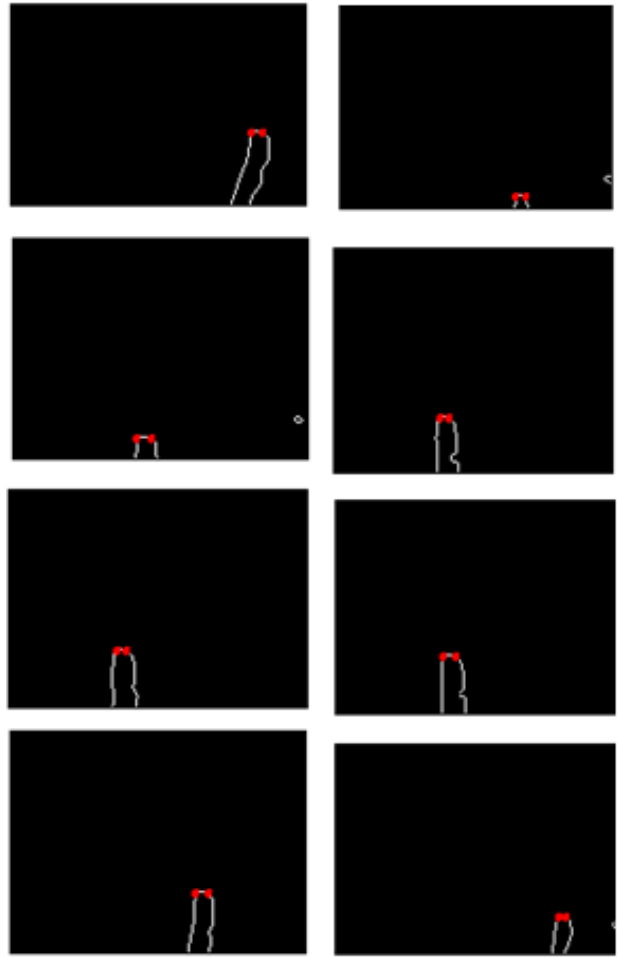

Figure 4: Highlighting the fingertips

All fingertips' pixel tracked in the process, were stored of that session and then showed on the screen. A movement of finger in one session is shown in figure 5. All pixels on the screen were connected by lines, in the same order, in which finger was moved. The connected lines are shown in figure 6. These lines will form the image on the screen which user drawing in air.

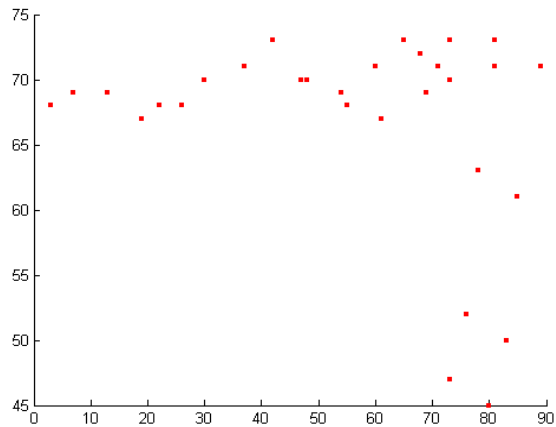

Figure 5: Displaying fingertips tracked on Screen in one session

In Figure 5 and 6 the screen positions are shown. The X and Y axis are mapped to the resolution of monitor screen.

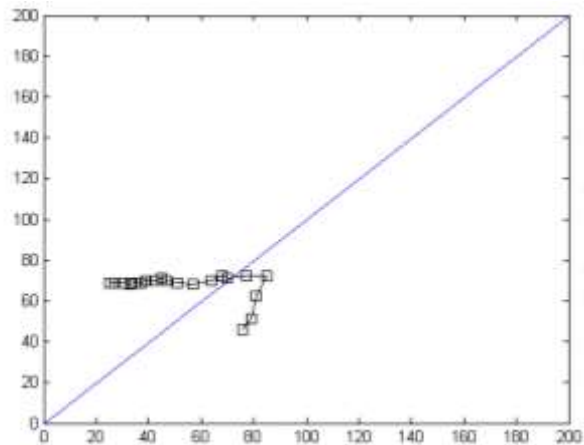

Figure 6: Gesture made by the user on Screen

The current methods for interaction are different where they use glove or sensors. So a direct comparison of our work is not possible, although the method used in this work can be compared with existing implemented methods. Bragatto [5] proposed method for sign language, which works well with recognition rate of 99.2 %. Araga [20] shows the accuracy of 99.0% for 5 different hand postures whereas it obtained accuracy of 94.3% for 9 gestures. Touch enable system where stylus or human fingers are used, have 100% accuracy of gesture recognition [10] [23-24]. A comparison of available methods are shown in given Table 1.

With the presented method, we are getting around 100% accuracy with simple backgrounds and 96% correct results on complex backgrounds. To make finger-stylus segmentation to be light intensity invariant, we tried to implement several methods which are discussed in [26-27]. The system works fine with varying light intensity; although a minimum threshold of light intensity is needed. The system takes 116 ms to show the first print of finger on screen and after that it continuously draws as the system is real time. The system was implemented with Matlab on Windows PC and webcam was capturing images as 12fps.

Also this system was implemented with MS Kinect where center of palm was used to draw on screen as shown in figure 7. Kinect based system work very same as the proposed one as Kinect recognized finger based on depth and any open finger can be recognized and traced. If user show any stick, not his finger, then also it would be detected as finger. This false detection scenario gives an upper hand to vision based approach than sensor based. Kinect based system gives 100% correct results with segmentation [28] but as Kinect is costlier in comparison of webcam and also it is a sensor, so we stick to webcam for final results.

Table 1: Comparison of Accuracy and Related Cost

| S. No. | Interaction Method | Accuracy | Devices needed and Issues |
|---|---|---|---|
| 1 | Gloves | 100% | Wired sensor gloves and User has to wear it ways. Its like using mouse. Sensor life is short. |
| 2 | Touch enable system [10] | 100% | Stylus and touch screen, Life is short. |
| 3 | Heat touch enable system [tablets] | 100% | Finger works, but touch screen is costly. |
| 4 | Hand Gesture [5][20][26] | 90-96% | Webcam, depend on image background. 100% on plain background. Cheaper than sensor, life is longer. |

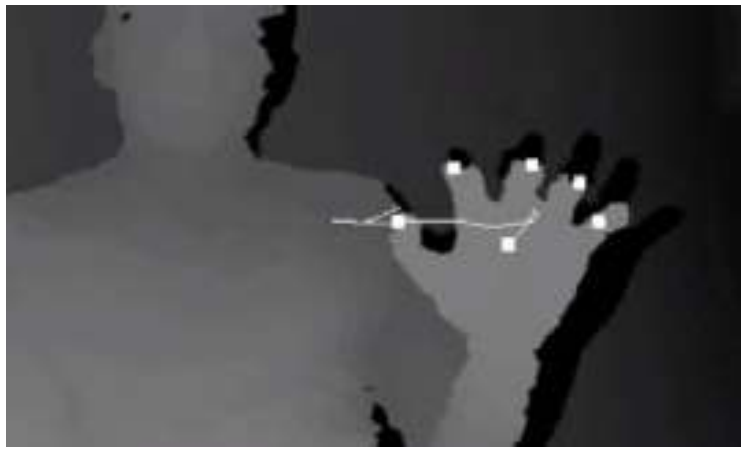

Figure 7: Results with MS Kinect

## V. CONCLUSION

Currently paint drawing technology uses touch enable screens or sensor gloves. In this paper natural-bare fingertip is used to draw visible content on screen in real time. Its like using stylus, where hand finger is doing that job. Many existing systems are available for commercial and academic purpose like *Sensu* stylus [10], games [24], touch enable drawing [11][23][29] etc use sensor or touch screens but a vision based paint drawing where no sensor is used, would be an advantage to reduce hardware cost and easy to use.

The proposed method provides a natural human-system interaction in such a way that it do not require keypad, stylus, digital pen or glove etc. for input. The method also shows the shape made by user virtually. The shape would be shown on screen which user has drawn in one session. This paper is an attempt to replace stylus with natural finger 'finger-stylus', where no need to touch the screen. In future, we would try to recognize these virtual shapes to bring them more useful recognitions and converting them into 3D for advanced applications. The system setup is shown in figure 8.

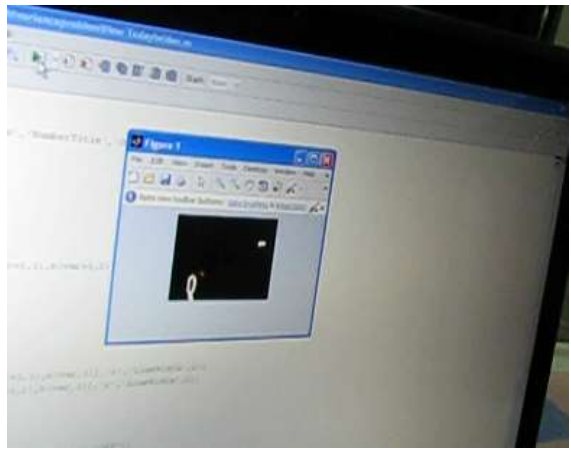

Figure 8: System setup